\documentclass[reprint,superscriptaddress,nofootinbib,amsmath,amssymb,floatfix,float,aps]{revtex4-2}
\usepackage{graphicx}
\usepackage{dcolumn}
\usepackage{bm}
\usepackage[colorlinks=true, allcolors=blue]{hyperref}
\usepackage{cleveref}
\usepackage{xcolor}
\usepackage[normalem]{ulem}

\NewDocumentCommand\mfs{O{}m}{{\color{red} \sout{#1}}~{\color{red} #2}}

\begin{document}

\title[]{A detailed first-order post-Newtonian analysis of massive Brans-Dicke theories: numerical constraints and the $\beta$ parameter meaning}

\author{Matheus F. S. Alves}
 \email{matheus.s.alves@edu.ufes.br}
\affiliation{Departamento de Física, Universidade Federal do Espírito Santo, Vitória, ES,  29075-910, Brazil.}
\affiliation{Núcleo de Astrofísica e Cosmologia (Cosmo-Ufes), Universidade Federal do Espírito Santo, Vitória, ES,  29075-910, Brazil.}

\author{Júnior D. Toniato}
 \email{junior.toniato@ufes.br}
\affiliation{Departamento de Química e Física, Universidade Federal do Espírito Santo - Campus Alegre, ES, 29500-000, Brazil.}%
\affiliation{Núcleo de Astrofísica e Cosmologia (Cosmo-Ufes), Universidade Federal do Espírito Santo, Vitória, ES,  29075-910, Brazil.}

\author{Davi C. Rodrigues}%
 \email{davi.rodrigues@ufes.br}
\affiliation{Departamento de Física, Universidade Federal do Espírito Santo, Vitória, ES,  29075-910, Brazil.}
\affiliation{Núcleo de Astrofísica e Cosmologia (Cosmo-Ufes), Universidade Federal do Espírito Santo, Vitória, ES,  29075-910, Brazil.}
\affiliation{Institute for Theoretical Physics, Heidelberg University, Philosophenweg 16, 69120 Heidelberg, Germany.}

\begin{abstract}
Massive Brans-Dicke (BD) theory is among the simplest general relativity extensions. It is commonly found as the weak-field limit of other gravitational theories.
Here we do a detailed post-Newtonian analysis of massive BD theories. 
We start by expanding the massive BD field equations following the Will-Nodtvedt Parameterized-Post-Newtonian (PPN) formalism, without point-particle approximations. 
A single potential that is not present in the standard PPN formalism is found. 
This new potential hinders immediate PPN conclusions. 
To proceed, we do a complete first-order post-Newtonian analysis and explicitly derive all the conserved quantities. 
After demanding that there exists a Newtonian limit by requiring the BD mass to be sufficiently large, we find, as expected, that $\gamma  = 1$; but there is no effective $\beta$ parameter that can have the same physical role of the standard $\beta$ in PPN formalism. All the others standard PPN parameters can be extended to the massive BD  case without issues and are shown to have the same values of general relativity.  
At last, we consider numerical relations on the periastron advance and the BD mass in two different physical contexts, the orbit of Mercury about the Sun and the orbit of the star S2 about the expected supermassive black hole in the Milky Way.
\end{abstract}

\maketitle

\section{INTRODUCTION}

Scalar-tensor theories are among the most popular alternative theories of gravity. Part of this status is due to the fact that adding to the metric tensor a scalar field to describe gravitational interaction can be seen as one of the most simple ways to modify general relativity without incour into higher than second-order field equations. The Brans-Dicke theory \cite{PhysRev.124.925}, the precursor of the scalar-tensor models, has a clear meaning for the scalar field $\Phi$: to replace the gravitational constant $G$ by a scalar function of the coordinates. The consequences of this theory are well known and exhaustively discussed in the literature \cite{bergmann1968comments,wagoner1970scalar,nordtvedt1970post,damour1992tensor,fujii2003scalar,faraoni2004scalar, PhysRevD.63.063504,EFlanagan_2004,clifton2012modified}.

A more sophisticated model emerges if one introduces to the original Brans-Dicke theory a potential function $V(\Phi)$ to drive its self-interactions \cite{santos1997cosmology}. The scalar-field dynamics assume the form of a massive field equation, from which these models are usually called as massive Brans-Dicke theories. The characteristic feature of these theories is the presence of a Yukawa-like potential in the weak-field approximation, breaking the Newtonian limit in general. 
The Newtonian limit can be restored if the scalar-field mass is either sufficiently small or sufficiently large (e.g., \cite{PhysRevD.104.044020}).

The breaking of the Newtonian limit makes not so simple the task of testing massive Brans-Dicke theories of gravity. Specifically, the theory is not suitable to the parametrized post-Newtonian formalism (PPN) \cite{carmichael1925eddington,robertson1962relativity, schiff1966comparison,PhysRev.169.1017,poisson_will_2014,Will_2014,will_2018}, a practical framework to compare theoretical predictions with solar system observational constraints \cite{fomalont2009progress,bertotti2003test,hofmann2010lunar,lambert2011improved, fienga2011inpop10a,pitjeva2013relativistic,Park_2017}. Although many conclusions has been drawn from the perspective of effective $\beta$ and $\gamma$ parameters (they are replaced by coordinate functions), we do not agree with such approach \cite{PhysRevD.81.047501,PhysRevD.94.124015,PhysRevD.88.084054}. The presence of a single non-standard potential to the metric expansion of the PPN formalism can be enough to modify the physical meaning of the PPN parameters. In this sense, to encode the PPN deviations into effective parameters that can vary in time and space does not guarantee that the observational constraint originally derived from the PPN formalism can still be applied to its effective counterpart.

A more safe procedure is to consider equations of motion and the definition of the conserved quantities, within the PPN approximation method, in order to infer how the new potentials affect the physical meaning of each PPN parameter. This was the methodology used to study Palatini $f(R)$ gravity in Ref. \cite{PhysRevD.101.064050}. For the specific case of scalar-tensor theories, Ref. \cite{PhysRevD.104.044020} made a detailed analysis of the $\gamma$ parameter, which is determined in the linear order of approximation. Moving to the non-linear order, a variety of new potentials arise, making this task considerably laborious.

However, once the use of PPN formalism presumes the existence of a Newtonian limit for the theory, it is reasonable for the scalar-tensor case to deal with a scalar field of negligible or large mass compared with the system scale. In the first scenario, the theory reduces to the original Brans-Dicke model, which post-Newtonian limit is well known. But, for the large mass case, the Yukawa correction can be transferred from the linear to the next order of approximation, leaving it as the single one potential not included in the PPN formalism, a way more treatable situation. 

This is the goal of this paper: to study the PPN parameters in a massive Brans-Dicke theory with a Yukawa-like potential at the non-linear order of approximation. The article is organized as follows. In Sec. \ref{sec:fieldeq} we describe and obtain the field equations of the Brans-Dicke theory with a potential. In Sec. \ref{sec:pn} we perform a post-Newtonian expansion of these field equations. In Sec. \ref{sec:conserved} we perform a derivation of all conserved quantities. In Sec. \ref{sec:eom} we proceed to obtain the acceleration of the center of mass. In Sec. \ref{sec:periastron} we discuss the periastron shift and the effect of the Yukawa correction in this phenomenon. In Sec. \ref{sec:beta} we discuss the meaning of the $\beta$ parameter and the Nordtvedt effect. We end with a conclusion in Sec. \ref{sec:conclusion}.

\section{MASSIVE BRANS-DICKE THEORIES} \label{sec:fieldeq}

Brans-Dicke theories are a variation of scalar-tensor theories of gravity, in which the gravitational constant $G$ is substituted by a scalar field $\Phi$, referred to as the Brans-Dicke field. The original theory only has a kinetic term that depends on a dimensionless function $\omega(\Phi)$ (in particular it can be a constant). 
Here we consider a Brans-Dicke theory with a potential $V(\Phi)$. We study this theory in a Post-Netonian context, aiming to find the Parametrized Post-Newotian (PPN) parameters when possible \cite{poisson_will_2014}.
In this context, and assuming a sufficiently smooth potential $V$, $V$ can be expanded in powers of $\Phi$. The relevant contribution from $V$ is the one that defines the Brans-Dicke field mass. These topics are further detailed below.

The Brans-Dicke action with a potential is given by
\begin{equation}
S=\int \frac{\sqrt{-g}}{2\kappa}\left[  \Phi R+2\frac{\omega\left(
\Phi\right)  }{\Phi}X-V\left(  \Phi\right)  \right]  d^{4}x+S_{m},
\label{action}%
\end{equation}
where $S_{m}$ is the action of the matter fields, $\kappa$ is the
dimensionless coupling constant, since we are using $G=1$, and%
\begin{equation}
X=-\frac{1}{2}g^{\mu\nu}\partial_{\mu}\Phi\partial_{\nu}\Phi,\label{defX}%
\end{equation}
is the kinetic term. There are two free functions in the theory, the scalar field potential $V$ and the coupling function $\omega$. We remark in this form Brans-Dicke theory is more general than metric and Palatini $f(R)$ models. Indeed, $\omega=0$ leads to the former \cite{Capone_2010,chiba20031,faraoni2007sitter}, while  $\omega=-3/2$ leads to the latter \cite{flanagan2004palatini,RevModPhys.82.451}.

The field equations are obtained from (\ref{action}) by varying it with respect
to $g^{\mu\nu}$ and $\Phi$:%
\begin{align}
G_{\mu\nu} &  =\frac{\kappa}{\Phi}T_{\mu\nu}+\frac{1}{\Phi}\left[  \nabla
_{\nu}\nabla_{\mu}\Phi-g_{\mu\nu}\square\Phi\right]  \label{fieldeq1}\\
&  +\frac{\omega}{\Phi^{2}}\left[  \partial_{\mu}\Phi\partial_{\nu}%
\Phi+Xg_{\mu\nu}-g_{\mu\nu}\frac{V}{2\Phi}\right]  ,\nonumber
\end{align}%
\begin{equation}
\frac{2\omega}{\Phi}\square\Phi=-R+\frac{2\omega^{\prime}}{\Phi}%
X-\frac{2\omega}{\Phi^{2}}X+V^{\prime},\label{fieldeq2}%
\end{equation}
where $G_{\mu\nu}$ is the Einstein tensor, $T_{\mu\nu}$ is the usual
energy-momentum tensor, $\nabla_{\mu}$ indicates a covariant derivative,
$\square\equiv\nabla_{\mu}\nabla^{\mu}$ is the d'Alembertian operator and the
prime ``${\;}^{\prime}$ " represents a derivative with respect to $\Phi$.

In the following section, we will expand these field equations in a post-Newtonian approximation.

\section{POST-NEWTONIAN LIMIT OF MASSIVE BRANS-DICKE THEORIES}
\label{sec:pn}

In this section, we will expand the field equations (\ref{fieldeq1}) and (\ref{fieldeq2}) up to the first post-Newtonian order. For this purpose, we make use of the PPN formalism following \cite{poisson_will_2014,will_2018,Will_2014}.

The starting point of our calculation is to consider the energy-momentum tensor as that of a perfect fluid, which is given by%
\begin{equation}
T^{\mu\nu}=\left(  \rho+\rho\Pi+p\right)  u^{\mu}u^{\nu}+pg^{\mu\nu},
\end{equation}
where $\rho$ is the mass density (we are working also with $c=1$), $\Pi$ is the fluid's internal energy per
unit mass, $p$ is the pressure and $u^{\mu}=u^{0}\left(  1,\mathbf{v}\right)
$ is the fluid four-velocity. Following the slow-motion condition, the
energy-momentum tensor components can be expanded in ``orders of smallness''
\begin{equation}
U\sim v^{2}\sim\frac{p}{\rho}\sim\Pi\sim O\left(  2\right) \, .
\end{equation}
As standard within PPN formalism, we use the notation $O\left(  N\right)  $ to represent quantities of order $v^{N}$ or smaller. Time derivatives are also taken to have an order of smallness associated with them, relative to spatial derivatives, i.e. $\partial_{0}\sim O\left(  1\right) $.

The fluid dynamics is subjected not only to null divergence of the
energy-momentum $\nabla^{\mu}T_{\mu\nu}=0$, but also to the conservation of
rest mass density,%
\begin{equation}
\nabla_{\mu}\left(  \rho u^{\mu}\right)  =0.
\end{equation}
This equation can be re-expressed as an effective flat-space continuity
equation as follows,%
\begin{equation}
\partial_{t}\left(  \rho^{\ast}\right)  +\partial_{i}\left(  v^{i}\right)  =0,
\label{conservation}%
\end{equation}
with%
\begin{equation}
\rho^{\ast}\equiv\sqrt{-g}u^{0}\rho. \label{rhoStar}%
\end{equation}
Latin sub-indexes are used for spatial components only.
From (\ref{conservation}) we see that the conserved mass density is $\rho^{\ast}$
and therefore it is more convenient to use it to express the energy-momentum
tensor components.

To accurately depict the motion of massive bodies in the first post-Newtonian approximation, it is sufficient to calculate $h_{00}$ up to $O\left(  4\right)  $, $h_{0i}$ up to $O\left(  3\right)  $, and $h_{ij}$ up to $O\left(  2\right)  $.

In order to utilize the PPN formalism in Brans-Dicke's theory, it is necessary to expand the scalar field, $\Phi$, in a post-Newtonian approximation. To achieve this, we expand $\Phi$ about a constant background given by $\varphi_0$. The latter is the value of $\Phi$ if the system in question is removed,

\begin{equation}
\Phi=\varphi_{0}+\varphi\text{, with } {\varphi_0} \sim
O\left(  0\right)  .
\end{equation}
Furthermore, we also need to expand the functions $\omega\left(  \Phi\right)  $
and $V\left(  \Phi\right)  $ around $\varphi_{0}$. Expanding these functions
we get%
\begin{align}
\omega &  =\omega_{0}+\omega_{1}\varphi+\omega_{2}\varphi^2 + \ldots \label{eq11}\\ 
V  &  =V_{0}+V_{1}\varphi+V_{2}\varphi^{2}+\ldots 
\end{align}
where all expansion coefficients we assume to be of $O\left(  0\right)  $. Following the PPN formalism, we neglect any effective cosmological constant that could emerge by considering $V_{0}=0$. Furthermore, from (\ref{fieldeq2}) in the lowest velocity order $O\left(  0\right)$ we obtain $V_{1}=0$. 

\subsection{Second order}

We start by considering the leading term in (\ref{fieldeq2}). Taking into
account the above expansions, we have
\begin{equation}
\left(  \nabla^{2}-m_{\varphi}^{2}\right)  \varphi=-\frac{\kappa}{\left(
2\omega_{0}+3\right)  }\rho^{\ast}, \label{scalarEq}%
\end{equation}
with the mass term given by,%
\begin{equation}
m_{\varphi}^{2}\equiv\frac{2\varphi_{0}}{\left(  2\omega_{0}+3\right)  }V_{2}.
\end{equation}
Equation (\ref{scalarEq}) solution, with the boundary condition that $\varphi$
should approach zero far from the source, is a Yukawa potential, which is
expressed by,%
\begin{equation}
\varphi=\frac{\kappa}{4\pi\left(  2\omega_{0}+3\right)  }\int\frac{\rho^{\ast
}\left(  t,\mathbf{x}^{\prime}\right)  }{\left\vert \mathbf{x-x}^{\prime
}\right\vert }e^{-m_{\varphi}\left\vert \mathbf{x-x}^{\prime}\right\vert
}d^{3}x^{\prime}.
\label{scalarYukawa}
\end{equation}

Keeping only the second order terms in (\ref{fieldeq1}), knowing that
$u^{0}\approx1$, due to its normalization and that the relevant component of
the energy-momentum tensor is $T^{00}\approx\rho\approx\rho^{\ast}$ because of Eq.
(\ref{rhoStar}), we obtain the following equations for the temporal and spatial
components of the metric tensor%
\begin{equation}
\nabla^{2}h_{00}^{\left(  2\right)  }=-\frac{\kappa}{\varphi_{0}}\rho^{\ast
}+\frac{1}{\varphi_{0}}\nabla^{2}\varphi, \label{00eq}%
\end{equation}

\begin{equation}
\nabla^{2}h_{ij}^{\left(  2\right)  }=-\frac{1}{\varphi_{0}}\left(
\delta_{ij}\nabla^{2}\varphi+\delta_{ij}\kappa\rho^{\ast}\right).
\label{ijeq}%
\end{equation}
In order to solve these equations we need to fix a gauge for the
metric tensor. A useful choice for the class of scalar-tensor theories we
consider is given by \cite{nutku1969post}%
\begin{equation}
\partial_{k}h_{i}^{k}+\frac{1}{2}\partial_{i}h_{00}-\frac{1}{2}\partial
_{i}h_{k}^{k}=\frac{1}{\varphi_{0}}\partial_{i}\varphi. \label{gaugechoice}
\end{equation}
In this gauge we can write the solution for (\ref{00eq}) and (\ref{ijeq}) as
\begin{equation}
h_{00}^{\left(  2\right)  }=\left(  \frac{\kappa}{4\pi\varphi_{0}}U+\frac
{1}{\varphi_{0}}\varphi\right)  ,
\end{equation}%
\begin{equation}
h_{ij}^{\left(  2\right)  }=\left(  \frac{\kappa}{4\pi\varphi_{0}}U-\frac
{1}{\varphi_{0}}\varphi\right)  \delta_{ij},
\end{equation}
where we are using
\begin{equation}
U=\int\frac{\rho^{\ast}\left(  t,\mathbf{x}^{\prime}\right)  }{\left\vert
\mathbf{x-x}^{\prime}\right\vert }d^{3}x^{\prime}.
\end{equation}

As discussed in Ref. \cite{PhysRevD.104.044020}, this theory has a Newtonian limit only if the mass is sufficiently small or sufficiently large.
In the first case, the exponential of eq. (\ref{scalarYukawa}) can be expanded and one finds that the metric assumes the standard PPN form, leading to the massless Brans-Dicke theory (e.g. \cite{klimek2009parameterized}). {The sufficiently large mass case is commonly evoked to vanishe the scalar field contribution, recovering general relativity results.
We recall that one of the assumptions of the PPN formalism is that any non-Newtonian correction must appear beyond the Newtonian order, never in the Newtonian order. An alternative for these cases is to perform a post-Yukawa expansion as discussed in Ref. \cite{PhysRevD.85.064041}.}\\

Here we are concerned with the scenario in which $m_\varphi$ is large enough to not violate the Newtonian limit only.
Thus, a natural assumption, in order to be consistent with standard post-Newtonian analysis, is to consider a scalar field mass sufficiently large such that any correction to the Newtonian physics appears only in the fourth order. Under such conditions, the Yukawa correction magnitude cannot be larger than $O(4)$, that is, it needs to satisfy,

\begin{equation}\label{cond}
\varphi\sim  O(4).
\end{equation}

Thus, from now on we restrict our analysis to models where the scalar field does not have impact on the Newtonian limit, but it will have influence on the first post-Newtonian order.\footnote{
The case $\varphi\sim O(3)$ cannot be considered within the PPN formalism, since 
the $g_{00}$ component is broken into $O(0)$, $O(2)$ and $O(4)$ terms. Therefore, a third-order scalar field would in general lead
to high-precision corrections on the Newtonian order that would also be
too large in comparison with the post-Newtonian corrections.
} 
Under the former assumption, the $O(2)$ terms are
\begin{equation}
h_{00}^{\left(  2\right)  }=\frac{\kappa}{4\pi\varphi_{0}}U\text{ \ and
\ }h_{ij}^{\left(  2\right)  }=\frac{\kappa}{4\pi\varphi_{0}}U\delta
_{ij}\text{.} \label{eq22}
\end{equation}

Furthermore, to obtain a well-posed Newtonian limit we must impose the
relation $\kappa/4\pi\varphi_{0}=2$. Therefore, the components of the
metric up to $O(2)$ are simply
\begin{equation}
h_{00}^{\left(  2\right)  }=2U,
\end{equation}%
\begin{equation}
h_{ij}^{\left(  2\right)  }=2U\delta_{ij}.
\end{equation} 
From the above, one concludes that the
second-order metric in this massive Brans-Dicke theory with $\varphi\sim O(4)$ [cf. Eq. \eqref{cond}] is identical as in GR. Once post-Newtonian corrections to the light motion are derived from this second-order metric only, one can also conclude that it necessarily yields 
\begin{equation}
\gamma = 1.    
\end{equation}
Therefore, the present model is consistent with observational data coming from light bending and Shapiro time delay. This is consistent with several works and a more precise discussion on the PPN parameter $\gamma$ in scalar-tensor theories can be found in Ref. 
\cite{PhysRevD.104.044020}.

\subsection{Third order}

Moving to the third order, only the $0j$-components of the field equation
(\ref{fieldeq1}) contribute, then we have the equation
\begin{equation}
\nabla^{2}h_{0j}^{\left(  3\right)  }=2\frac{\kappa}{\varphi_{0}}v_{j}%
\rho^{\ast}-\frac{1}{2}\partial_{0}\partial_{j}h_{00}^{\left(  2\right)  }.
\end{equation}
The solution is then easily obtained to be%
\begin{equation}
h_{0j}^{\left(  3\right)  }=-4U_{j}-\frac{1}{2}\partial_{tj}X, \label{eq26}
\end{equation}
where
\begin{equation}
U_{j}=\int\frac{\rho^{\ast}\left(  t,\mathbf{x}^{\prime}\right)  v_{j}%
}{\left\vert \mathbf{x}-\mathbf{x}^{\prime}\right\vert }d^{3}x^{\prime},
\end{equation}
and%
\begin{equation}
X=\int\rho^{\ast}\left(  t,\mathbf{x}^{\prime}\right)  \left\vert
\mathbf{x}-\mathbf{x}^{\prime}\right\vert d^{3}x^{\prime}.
\end{equation}

\subsection{Fourth order}

To derive the $00$-component of the metric in $O(4)$ we first compute $u^{0}$
and $\rho^{\ast}$ up to $O\left(  2\right)  $,%
\begin{equation}
u^{0}=1+\left(  2U+v^{2}\right)  ,
\end{equation}%
\begin{equation}
\rho^{\ast}=\rho\left(  1+3U+\frac{v^{2}}{2}\right)  .
\end{equation}
With the expressions above, we find
\begin{equation}
\nabla^{2}h_{00}^{\left(  4\right)  }=\nabla^{2}\left[  3\Phi_{1}-2\Phi
_{2}+2\Phi_{3}+6\Phi_{4}-2U^{2}+\varphi\right]  . \label{31}
\end{equation}
Therefore, by using that the metric is asymptotically flat, the
general solution reads
\begin{equation}
h_{00}^{\left(  4\right)  }=3\Phi_{1}-2\Phi_{2}+2\Phi_{3}+6\Phi_{4}%
-2U^{2}+a_{1}\Phi_{Y},
\end{equation}
where%
\begin{equation}
\Phi_{1}=\int\frac{\rho^{\ast\prime}v^{\prime2}}{\left\vert \mathbf{x}%
-\mathbf{x}^{\prime}\right\vert }d^{3}x^{\prime}\text{, \ }\Phi_{2}=\int
\frac{\rho^{\ast\prime}U^{\prime}}{\left\vert \mathbf{x}-\mathbf{x}^{\prime
}\right\vert }d^{3}x^{\prime}\text{,\ }%
\end{equation}%
\begin{equation}
\Phi_{3}=\int\frac{\rho^{\ast\prime}\Pi^{\prime}}{\left\vert \mathbf{x}%
-\mathbf{x}^{\prime}\right\vert }d^{3}x^{\prime}\text{, \ }\Phi_{4}=\int
\frac{p^{\prime}}{\left\vert \mathbf{x}-\mathbf{x}^{\prime}\right\vert }%
d^{3}x^{\prime}\text{,}%
\end{equation}%
\begin{equation}
\Phi_{Y}=\int\frac{\rho^{\ast\prime}}{\left\vert \mathbf{x-x}^{\prime
}\right\vert }e^{-m_{\varphi}\left\vert \mathbf{x-x}^{\prime}\right\vert
}d^{3}x^{\prime}\text{,} \label{poty}
\end{equation}
and%
\begin{equation}
a_{1}=\frac{2\varphi_{0}}{\left(  2\omega_{0}+3\right)  }\text{.}%
\end{equation}
The primed fluid variables are evaluated at time $t$ and position $\mathbf{x^{\prime}}$.

Collecting together the previous results, one can write
\begin{equation}
g_{00}=-1+2U+2\left(  \psi-U^{2}\right)  +a_{1}\Phi_{Y}, \label{metric1}%
\end{equation}%
\begin{equation}
g_{0j}=-4U_{j}-\frac{1}{2}\partial_{tj}X, \label{metric2}%
\end{equation}%
\begin{equation}
g_{ij}=\left(  1+2U\right)  \delta_{ij}, \label{metric3}%
\end{equation}
where we have defined,
\begin{equation}
\psi\equiv\frac{3}{2}\Phi_{1}-\Phi_{2}+\Phi_{3}+3\Phi_{4}.
\end{equation}

As expected, the limit of general relativity is obtained when $\omega
_{0}\rightarrow\infty$ with $\varphi_0$ fixed. Under some specific situations,  GR may not be recovered in the previous limit (e.g, \cite{faraoni1998conformal}), indeed, when $\varphi_0$ depends on $\omega_0$.

In the metric above, we have the presence of a new potential $\Phi_{Y}$ that is outside the standard Will-Nordvedt parameterization, so it is not correct to infer any immediate limit on any PPN parameter.

To obtain information about the other parameters it is necessary to derive the
equations of motion of the fields of matter to confront the theory with
observational data from experiments performed in the solar system. For this, a few more
details are needed and the first step is to examine the PN hydrodynamics in
order to find the conserved quantities.

\section{CONSERVED QUANTITIES}
\label{sec:conserved}

Within the PPN formalism, five of its ten parameters are directly related to
the possibility of a theory to satisfy conservation laws for the total energy
and total momentum \cite{poisson_will_2014,will_2018,Will_2014}. These laws can be obtained from the integration of the
energy-momentum tensor conservation equation 
\begin{equation}
    \nabla_{\mu}T^{\mu\nu}=0 \, ,
\end{equation}
in a finite-volume $V$ with boundaries outside the region occupied by matter.

The re-scaled density $\rho^{\ast}$ defined in (\ref{rhoStar}) allows us to
defined the fluid total rest mass in a given volume $V$ as,%
\begin{equation}
m=\int_{V}\rho^{\ast}d^{3}x. \label{mass}%
\end{equation}
Using that for any arbitrary function $f(\textbf{x},t)$, one has 
\begin{equation}
\frac{d}{dt}\int_{V}\rho^{\ast}f\left(  t,\mathbf{x}\right)  d^{3}x=\int_{V}\rho^{\ast}\frac{df\left(  t,\mathbf{x}\right)  }{dt}d^{3}x \label{prop}, 
\end{equation}
we have that the mass $m$ is conserved.
The steps to obtain (\ref{prop}) are straightforward, and they rely on the continuity equation (\ref{conservation}), Gauss's theorem, and the fact that $\rho^{\ast}$ vanishes on the boundary of the domain of integration. 

From the energy-momentum tensor conservation, we write
\begin{equation}
\partial_{\nu}\left(  \sqrt{-g}T^{\mu\nu}\right)  +\Gamma_{\alpha\nu}^{\mu
}\left(  \sqrt{-g}T^{\alpha\nu}\right)  =0.
\label{energyMomentumEq}%
\end{equation}
For $\mu=0$ and up to $O(5)$, one gets%
\begin{equation}
\rho^{\ast}\frac{d}{dt}\left(  \frac{1}{2}v^{2}+\Pi\right)  +\partial
_{j}\left(  pv^{j}\right)  -\rho^{\ast}v^{j}\partial_{j}U=0.
\end{equation}
This result can be expressed as an energy conservation statement for a bounded
matter source, i.e.
\begin{equation}
\frac{dE}{dt}=0,
\end{equation}
with
\begin{equation}
E=\int\left(  \frac{1}{2}\rho^{\ast}v^{2}+\rho^{\ast}\Pi-\frac{1}{2}\rho
^{\ast}U\right)  d^{3}x. \label{Energy}%
\end{equation}
The total mass-energy of the fluid is defined as,%
\begin{equation}
M=m+E, \label{mass-energyDef}%
\end{equation}
and, through (\ref{mass}) and (\ref{Energy}), it satisfies $dM/dt=0$.

Considering now $\mu=i$ up to $O(6)$ one finds
\begin{gather}
\partial_{0}\left(  \mu\rho^{\ast}v_{j}\right)  +\partial_{k}\left(  \mu
\rho^{\ast}v^{j}v^{k}\right)  +\partial_{j}p-\rho^{\ast}\partial
_{j}U\label{MometumO6}\\
-\rho^{\ast}\left(  \frac{3}{2}v^{2}-U+\Pi+\frac{p}{\rho^{\ast}}\right)
\partial_{j}U-\rho^{\ast}\partial_{j}\Psi\nonumber\\
+2\rho^{\ast}\frac{d}{dt}\left(  Uv_{j}\right)  +2U\partial_{j}p-4\rho^{\ast
}\frac{dU_{j}}{dt}+4\rho^{\ast}v^{k}\partial_{j}U_{k}\nonumber\\
-\frac{1}{2}\rho^{\ast}a_{1}\partial_{j}\Phi_{Y}=0,\nonumber
\end{gather}
with
\begin{equation}
\mu\equiv1+\frac{1}{2}v^{2}+U+\Pi+\frac{p}{\rho^{\ast}}.
\end{equation}
We next integrate (\ref{MometumO6}) over the volume occupied by the fluid. All
terms with the exception of the term with $\Phi_{Y}$ \ are general relativity
terms and can be found in \cite{poisson_will_2014}. To
integrate the extra term we use the \textquotedblleft switch
trick\textquotedblright, which consists in interchanging the variables
$\mathbf{x}\leftrightarrow\mathbf{x}^{\prime}$ inside the integral. This leads
us to
\begin{equation}
\int\rho^{\ast}\partial_{j}\Phi_{Y}d^{3}x=0.
\end{equation}

At the end, one finds the following vector conservation law,
\begin{equation}
\frac{dP_{j}}{dt}=0,
\end{equation}
with%

\begin{gather}
P_{j}=\int\rho^{\ast}v_{j}\left(  1+\frac{1}{2}v^{2}-\frac{1}{2}U+\Pi+\frac
{p}{\rho^{\ast}}\right)  d^{3}x\label{Momentum}\\
-\frac{1}{2}\int\rho^{\ast}\Phi_{j}d^{3}x.\nonumber
\end{gather}

From the above one sees that $P_j$ does not depend on $\Phi_Y$ and that it has exactly the same expression as in GR.

The previous results show that massive Brans-Dicke theories $\varphi\sim O(4)$, do not violate the total conservation of energy and momentum up to first post-Newtonian order. This is a direct consequence of all the PPN parameters $\zeta$'s and $\alpha_3$ being null, i.e.
\begin{equation}
\zeta_{1}=\zeta
_{2}=\zeta_{3}=\zeta_{4}=\alpha_{3}=0.    
\end{equation}
Therefore, equations (\ref{Energy}) and (\ref{Momentum}) can be directly compared to their counterparts in the PPN formalism \cite{will_2018}. For $E$ the expression is identical to \eqref{Energy}, not being dependent on any PPN parameter. For the momentum one has
\begin{align}
    P_j^{\rm PPN}=& \int\rho^{\ast}v_{j}\left(  1+\frac{1}{2}v^{2}-\frac{1}{2}U+\Pi+\frac
{p}{\rho^{\ast}}\right)  d^{3}x\label{Momentum-PPN}\\
&-\frac{1}{2}\int\rho^{\ast}\left[(1+\alpha_2)\Phi_{j}+\frac{1}{2}(\alpha_1-\alpha_2)V_{j}\right]d^{3}x,\nonumber
\end{align}
which gives
\begin{align}
    \alpha_1=\alpha_2=0,
\end{align}
for the massive Brans-Dicke theory with $\varphi\sim O(4)$. Thus, the model is a fully conservative one and it also does not present prefered-frame effects.

\section{EQUATION OF MOTION FOR MASSIVE BODIES}
\label{sec:eom}

In this section, we want to obtain the PN equations of motion for the
center of mass of massive bodies. For that, we split the fluid description of
the source into $N$ separated bodies. This is a realistic way to deal with the
trajectories of massive and finite-volume bodies, instead of assuming test
particles. Each body indexed by $A$ has a total rest mass given by%
\begin{equation}
m_{A}=\int_{A}\rho^{\ast}d^{3}x.
\end{equation}
The volume of integration above is calculated as a time-independent region of space that extends beyond the volume occupied by the body. Let's assume that this volume is large enough that, in a time interval $dt$, the body does not cross its boundary surface but it is also small enough to not intersect with any other body of the system. The center of mass, its velocity, and acceleration of a body $A$ are then defined as
\begin{equation}
\mathbf{r}_{A}\equiv\frac{1}{m_{A}}\int_{A}\rho^{\ast}\mathbf{x}d^{3}x,  
\end{equation}
\begin{equation}
 \mathbf{v}_{A}\equiv\frac{d\mathbf{r}_{A}}{dt}=\frac{1}{m_{A}}\int_{A}
\rho^{\ast}\mathbf{v}d^{3}x,   
\end{equation}
\begin{equation}
\mathbf{a}_{A}\equiv\frac{d\mathbf{v}_{A}}{dt}=\frac{1}{m_{A}}\int_{A}
\rho^{\ast}\frac{d\mathbf{v}}{dt}d^{3}x. \label{aEq}
\end{equation}

Following \cite{poisson_will_2014}, the center of mass acceleration of each body is decomposed into three parts,%
\begin{equation}
\mathbf{a}_{A}=\mathbf{a}_{A}\left[  \text{Newt}\right]  +\mathbf{a}%
_{A}\left[  \text{PN}\right]  +\mathbf{a}_{A}\left[  \text{Str}\right]  .
\end{equation}
The first term is the Newtonian contribution. The second term is the PN corrections apart from any contribution due to the internal structure of the bodies, which are encoded within the third term.
In order to obtain the integrand of eq. (\ref{aEq}) we need the PN extension of
Euler equation, which comes from (\ref{energyMomentumEq}), and it reads%

\begin{gather}
\rho^{\ast}\frac{dv^{j}}{dt}=-\partial_{j}p+\rho^{\ast}\partial_{j}%
U+\rho^{\ast}\left(  \frac{1}{2}v^{2}+U+\Pi+\frac{p}{\rho^{\ast}}\right)
\partial_{j}p\nonumber\\
-v^{j}\partial_{t}p+\rho^{\ast}\left(  v^{2}-4U\right)  \partial_{j}%
U-\rho^{\ast}v^{j}\left(  3\partial_{t}U+4v^{k}\partial_{k}U\right)
\nonumber\\
+4\rho^{\ast}\partial_{t}U_{j}+4\rho^{\ast}v^{k}\left(  \partial_{k}%
U_{j}-\partial_{j}U_{k}\right)  \nonumber\\
+\rho^{\ast}\partial_{j}\Psi+\frac{1}{2}\rho^{\ast}a_{1}\partial_{j}\Phi
_{Y}=0. \label{EulerEq}
\end{gather}

To obtain the acceleration of the center of mass we substitute (\ref{EulerEq})
in (\ref{aEq}) and calculate the integrals. As before, the only term different
from general relativity is the one containing the new potential $\Phi_{Y}%
$. To integrate this term we use the fact that the
gravitational potentials can be separated into an internal part, produced by body
$A$, and an external part originated by the other bodies of the system. When
integrating the terms containing the inner parts, both the integrals will have
the same domain, so they can be calculated with the help of the switch trick
mentioned earlier. Assuming a large separation between the bodies implies
that, when evaluating an external potentials inside the body $A$, they can be
expanded in a Taylor series. In this context we have%
\begin{equation}
\Phi_{Y,A}^{ext}\approx\Phi_{Y,A}^{ext}\left(  t,\mathbf{r}_{A}\right)
+\bar{x}^{j}\partial_{j}\Phi_{Y,A}^{ext}\left(  t,\mathbf{r}_{A}\right)  +...\,, \label{eq60}
\end{equation}
where $\mathbf{\bar{x}}$ gives the position of a fluid element relative to the center of mass $\mathbf{r_A}(t)$.
This expansion is used to extract the outer pieces of potentials from the
integrals and obtain
\begin{equation}
\int_{A}\rho^{\ast}\partial_{j}\Phi_{Y}d^{3}x=m_{A}\partial_{j}\Phi
_{Y,A}^{ext}.
\label{eq61}
\end{equation}
The above result is achieved because the seccond term in Eq.(\ref{eq60}) contains an odd number of internal vectors and, by considering that the bodies are ``reflect-symmetric'' with respect to their center of mass, i.e. $\rho^*(t,\mathbf{\bar{x}})=\rho^*(t,-\mathbf{\bar{x}})$, their integral vanishes.
With this result, we calculate the individual contribution of this new term to
the gravitational force. We get
\begin{equation}
\partial_{j}\Phi_{Y,A}^{ext}=-\sum_{B\neq A}\left(  1+m_{\varphi}%
r_{AB}\right)  \frac{m_{B}n_{AB}^{j}}{r_{AB}^{2}}e^{-m_{\varphi}r_{AB}}.
\end{equation}
\begin{widetext}
In the end, one finds the following center of mass acceleration
\begin{align}
\mathbf{a}_{A}\left[  \text{Newt}\right]  &  =-\sum_{B\neq A}\frac{m_{B}}%
{r_{AB}^{2}}\mathbf{n}_{AB},\label{accN}\\
\mathbf{a}_{A}\left[  \text{PN}\right]   &  =-\sum_{B\neq A}\frac{m_{B}%
}{r_{AB}^{2}}\left[  v_{A}^{2}-4\left(  \mathbf{v}_{A}\cdot\mathbf{v}%
_{B}\right)  +2v_{B}^{2}-\frac{3}{2}\left(  \mathbf{n}_{AB}\cdot\mathbf{v}%
_{B}\right)  ^{2}-\frac{5m_{A}}{r_{AB}}-\frac{4m_{B}}{r_{AB}}\right]
\mathbf{n}_{AB}\label{accPN}\\
&  +\sum_{B\neq A}\frac{m_{B}}{r_{AB}^{2}}\left[  \mathbf{n}_{AB}\cdot\left(
4\mathbf{v}_{A}-3\mathbf{v}_{B}\right)  \right]  \left(  \mathbf{v}%
_{A}-\mathbf{v}_{B}\right) \nonumber\\
&  +\sum_{B\neq A}\sum_{C\neq A,B}\frac{m_{B}m_{C}}{r_{AB}^{2}}\left[
\frac{4}{r_{AC}}+\frac{1}{r_{BC}}-\frac{r_{AB}}{2r_{BC}^{2}}\left(
\mathbf{n}_{AB}\cdot\mathbf{n}_{BC}\right)  \right]  \mathbf{n}_{AB}%
\nonumber\\
&  -\frac{7}{2}\sum_{B\neq A}\sum_{C\neq A,B}\frac{m_{B}m_{C}}{r_{AB}%
r_{BC}^{2}}\mathbf{n}_{BC}-\frac{1}{2}a_{1}\sum_{B\neq A}\left(  1+m_{\varphi
}r_{AB}\right)  \frac{m_{B}}{r_{AB}^{2}}e^{-m_{\varphi}r_{AB}}\mathbf{n}%
_{AB},\nonumber
\end{align}

\begin{equation}
\mathbf{a}_{A}\left[  \text{Str}\right]  =-\sum_{B\neq A}\frac{E_{B}}%
{r_{AB}^{2}}\mathbf{n}_{AB},\label{accStr}
\end{equation}
where
\begin{equation}
E_{B}=T_{B}+\Omega_{B}+E_{B}^{int},
\end{equation}
is the total energy. In the above expressions, we use the definitions
$\mathbf{r}_{AB}=\mathbf{r}_{A}-\mathbf{r}_{B}$, $r_{AB}=\left\vert
\mathbf{r}_{AB}\right\vert $ and $\mathbf{n}_{AB}=\frac{\mathbf{r}_{AB}%
}{r_{AB}}$.
\end{widetext}
Its is worth to note that $\mathbf{a}_{A}\left[  \text{Newt}\right]$ and $\mathbf{a}_{A}\left[  \text{Str}\right]$ together form the quasi-Newtonian contribution to the body's $A$ acceleration. It goes with the inverse square of the distance between bodies $A$ and $B$, but it is proportional to the total mass-energy of body $B$, namely $M_B$ [cf. Eq. \eqref{mass-energyDef}]. Moreover, the terms $m_B$ in $\mathbf{a}_{A}\left[\text{PN}\right]$ can be substituted by $M_B$ without affecting the dynamics within the first PN order. Therefore, we can conclude that the equations of motion depends on the bodies internal structure only through their total mass-energy. The independence with respect to any specific internal quantity, like the body's gravitational energy $\Omega_B$, is a confirmation that the theory satisfies the strong equivalence principle.

\section{PERIASTRON ADVANCE}
\label{sec:periastron}

The equation for the acceleration of the center of mass obtained earlier
applies to any number of well-separated bodies. Now let us specialize to a two-body system with the center of mass at the origin; to this end let us
define%
\begin{equation}
m:=M_{1}+M_{2},\text{ } \text{ } \eta:=\frac{M_{1}M_{2}}{\left(  M_{1}+M_{2}\right)
^{2}},
\end{equation}
where $m$ is a type of total mass and $\eta$ a symmetric mass ratio; it should be noted that
$m$ differs from the total mass-energy $M$ introduced earlier by terms of
order $O\left(  2\right)  $. Let's also introduce the separation
$\mathbf{r}:=\mathbf{r}_{1}-\mathbf{r}_{2}$, the relative velocity
$\mathbf{v}:=\mathbf{v}_{1}-\mathbf{v}_{2}$, and let's make $r:=\left\vert
\mathbf{r}\right\vert =r_{12}$, $\mathbf{n}:=\frac{\mathbf{r}}{r}%
=\mathbf{n}_{12}$ and $v:=\left\vert \mathbf{v}\right\vert $ . Writing
$\mathbf{\dot{r}}:=\mathbf{v}\cdot\mathbf{n}$ we have
\begin{align}
\mathbf{a}   =& -\frac{m}{r^{2}}\mathbf{n}-\frac{m}{r^{2}}\left\{  \left[
\left(  1+3\eta\right)  v^{2}-\frac{3}{2}\eta\dot{r}^{2}-2\left(
2+\eta\right)  \frac{m}{r}\right]  \mathbf{n}\right.  \notag\\
& \left.  -2\left(  2-\eta\right)  \dot{r}\mathbf{v}+\frac{a_{1}}{2}\left(
1+m_{\varphi}r\right)  e^{-m_{\varphi}r}\mathbf{n}\right\}  .\label{acc}
\end{align}
With the equation of motion in hand, we can visualize the post-Newtonian
corrections together with the terms originated by $\Phi_{Y}$ as perturbations
of Kepler's orbit and employ the method of osculating elements \cite{poisson_will_2014} to obtain an expression for the periastron
advance of the binary system.
\begin{widetext}
Using the method of osculating elements we arrive at the equations%

\begin{equation}
\frac{dp}{df}\simeq4m\left(  2-\eta\right)  e\sin f, \label{secularP}%
\end{equation}%
\begin{align}
\frac{de}{df}  \simeq & \frac{m}{p}\left[  \left(  3-\eta+\frac{1}{8}\left(
56-47\eta\right)  e^{2}\right)  \sin f+\left(  5-4\eta\right)  e\sin
2f-\frac{3}{8}\eta e^{2}\sin3f\right] \label{secularE}\\
&  -\frac{a_{1}}{2}\sin f\left(  1+m_{\varphi}r\right)  e^{-m_{\varphi}%
r},\nonumber
\end{align}%
\begin{align}
\frac{d\omega}{df}  \simeq & \frac{1}{e}\frac{m}{p}\left[  3e-\left(
3-\eta-\frac{1}{8}\left(  8+21\eta\right)  e^{2}\right)  \cos f-\left(
5-4\eta\right)  e\cos2f+\frac{3}{8}\eta e^{2}\cos3f\right] \label{secularW}\\
&  +\frac{a_{1}}{2e}\cos f\left(  1+m_{\varphi}r\right)  e^{-m_{\varphi}%
r}.\nonumber
\end{align} 
\end{widetext}
From equations (\ref{secularP}), (\ref{secularE}) and (\ref{secularW}) we can
calculate the secular change in the Keplerian orbital parameters, the semi-latus rectum $p$, the
longitude of pericenter $\omega$ and the eccentricity
$e$, all three produced by post-Newtonian perturbations. It is worth to note that variations on $p$ is not affected by the scalar field, and this is due to the fact that the Yukawa-like term in the relative acceleration \eqref{acc} has influence in the radial direction only. Once $p$ is directly related with the body's angular-momentum norm, only force terms acting in the angular direction can affect its variation.

In order to obtain the secular
changes we integrate the above equations over a complete orbital period (from
$f=0$ to $f=2\pi$) and one gets%
\begin{equation}
    \Delta p=0,
\end{equation}
\begin{equation}
\text{\ }\Delta e=0,
\end{equation}%
\begin{equation}
\Delta\omega= \frac{6\pi m}{p} + \delta\omega,
\label{periastronAdvance}%
\end{equation}
where,
\begin{equation}
    \delta\omega = \frac{a_{1}}{2e}\int_{0}^{2\pi}\left(
1+m_{\varphi}r\right)  \cos fe^{-m_{\varphi}r}df,
\label{int_exp}
\end{equation}
is the correction due to Yukawa's contribution. 
The fact that $p$ undergoes no secular change is a consequence of
angular-momentum conservation and the absence of a secular change in $e$ is a consequence of energy conservation. The only parameter that undergoes a
secular evolution is the longitude of periastron, $\omega$.

\begin{figure}[t]
\includegraphics[width=\columnwidth]{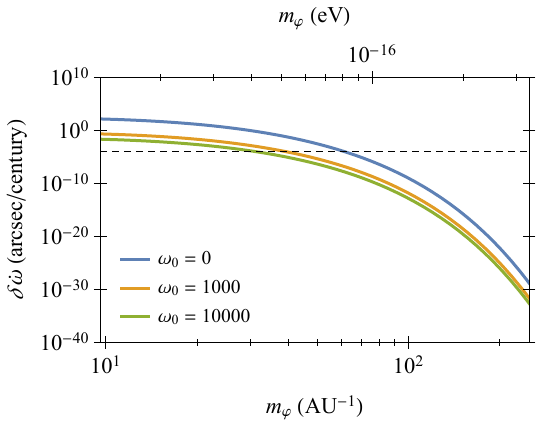}
\caption{Numerical solution for $\delta\dot\omega$ as a function of $m_\varphi$, in the case of Mercury ($e=0.2056$, $p=0.370$ AU, $P = 87.97$ days), is shown for different $\omega_0$ values. The black-dashed line is the uncertainty limit of Mercury's perihelion shift \cite{Park_2017}.}
\label{fig1}
\end{figure}

The advance per orbit $\Delta \omega$ can be converted to a rate by dividing by the orbital period, $P$. The result is
\begin{equation}
    \dot\omega = 3\left(\frac{2\pi}{P}\right)^{5/3}\frac{m^{2/3}}{1-e^2} \ + \ \delta\dot\omega,
    \label{eq77}
\end{equation}
after using Kepler's third law, $P^2=4\pi^2a^3/m$, with $a$ being the semi-major axis, defined by $p=a(1-e^2)$. The first term on the right hand side of \eqref{eq77} is the GR standard result, and $\delta\dot\omega=\delta\omega/P$ is the scalar-field contribution to the periastron secular variation.

One can verify that, as expected, for constant $r$ one finds $\delta \omega = 0$. In order to understand the physical impact of $\delta \omega$, we proceed with a numerical analysis. To illustrate the analysis, we consider Mercury's orbit. Since
\begin{equation}
r=\frac{p}{\left(  1+e\cos f\right)  }\label{r_f_1} \, ,
\end{equation}
we substitute (\ref{r_f_1}) into (\ref{int_exp}). Integrating the latter, we can get the curves for $\delta\dot\omega$ values shown in Figure \ref{fig1}.
The lower mass bound utilized in this analysis ensures the preservation of the Newtonian limit. For the solar system, this bound is determined by considering a reference distance of $r = 1$ AU and an approximation of $U \approx 10^{-8}$. By imposing the condition under which we are working, specifically $\varphi\sim  O(4)$, and a constant density approximation, we can infer that $m_\varphi\gtrsim 9.6 \ \text{AU}^{-1}$ or $m_\varphi\gtrsim 1.3\times10^{-17} \ \text{eV}$.

\begin{figure}[t]
\includegraphics[width=\columnwidth]{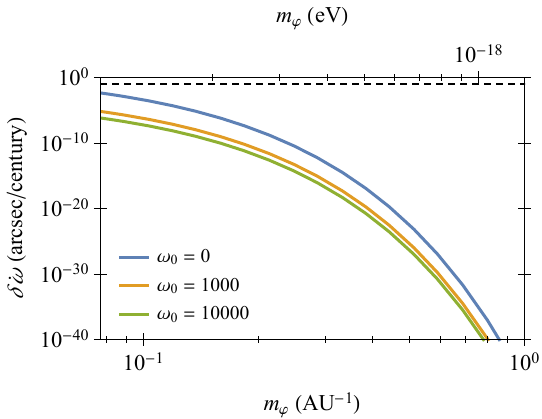}
\caption{Numerical solution for $\delta\dot\omega$ as a function of $m_\varphi$, in the case of S2 ($e=0.88$, $p=210.8$ AU, $P = 87.97$ days), is shown for different $\omega_0$ values. The black-dashed line is the uncertainty limit of S2 perihelion shift \cite{GRAVITY:2020gka}.}
\label{fig5}
\end{figure}

Our analysis is not limited to the solar system. As an example, we consider the star S2 orbiting  Sgr $\text{A}^\ast$, which is likely to be a supermassive black hole.  Here it is only assumed that S2 orbits a massive compact object, it needs not to be a black hole (i.e., a stationary compact massive object with a horizon \cite{Sotiriou:2011dz}).
To determine the lower mass bound in this case, we take the pericenter distance of S2 as the reference, denoted as $r = 120$ AU \cite{GRAVITY:2020gka}. Given that the orbital velocity $v$ is approximately $7700  \ \text{km} \ s^{-1}$, we have $U \sim v^2 \approx10^{-4}$. By applying our condition, considering a distribution of a point mass we obtain $m_{\varphi}\gtrsim 0.077 \ \text{AU}^{-1}$ or $m_{\varphi}\gtrsim 1.02 \times 10^{-19} \ \text{eV}$. Following that, we get the curve for $\delta\omega$ values shown in Figure \ref{fig5}. It is worth noting that the influence of the spin over the pericenter advance of S2 is smaller than $O(4)$, as has been discussed in references \cite{Merritt_2010,Zhang_2017}.

We are treating $\delta\dot\omega$ as the observational error in (\ref{eq77}) and checking for which values of $m_{\varphi}$ and $\omega_0$ (which is related to $\omega(\varphi)$ via Eq. (\ref{eq11})) the Yukawa potential $\Phi_Y$ would not influence the observations because $\delta\dot\omega$ would not be detected. Therefore, values of $\delta\dot\omega$ below the black dashed line, which corresponds to the uncertainty value of the experiment, imply values for $m_\varphi$ and $\omega_0$ in which this model has not observational influence of $\Phi_Y$. 
From figures \ref{fig1} and \ref{fig5} we can see that the Yukawa correction decreases as the mass of the scalar field increases. Furthermore, the lower the value of $\omega_0$, the more effective the correction, $\delta\dot\omega$, and consequently, we have a bigger deviation from the GR result.

Once the Yukawa correction (\ref{int_exp}) depends on the orbital parameters $p$ and $e$, we will study the influence of these parameters on $\delta\dot\omega$. First, in order to exemplify the effect of eccentricity $e$ in the correction $\delta\dot\omega$ we consider a semi-latus rectum of $p=0.3$ AU. 
\begin{figure}[t!]
\includegraphics[width=\columnwidth]{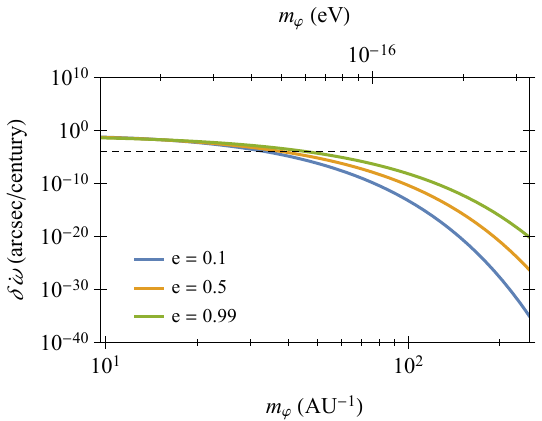}
\caption{Numerical solution for $\delta\dot\omega$. It was used $\omega_{0} = 4000$, $p=0.3$ AU and $P = 87.97$ days for the \textit{semi-latus rectum} and the orbital period. The blue, yellow, and green curves are constructed with $e=0.1$, $e=0.5$, and $e=0.99$, respectively. The black dashed line is the uncertainty of the Mercury perihelion shift experiment \cite{Park_2017}.}
\label{fig2}
\end{figure}
Figure \ref{fig2} shows that orbits with a bigger eccentricity have a greater effect of the Yukawa correction. This is an expected result since more eccentric orbits have a bigger periastron advance. To study the effect of $p$ on the correction $\delta\dot\omega$ we need to fix the eccentricity $e$. Note that as we do this the semi-latus rectum $p$ measures the average distance between the two bodies of our system. To exemplify this effect we consider the eccentricity of $e = 0.5$. As shown in Figure \ref{fig3}, the bigger the distance, the smaller the effect of the $\delta\dot\omega$ correction. This is also 
expected.

\begin{figure}[t!]
\includegraphics[width=\columnwidth]{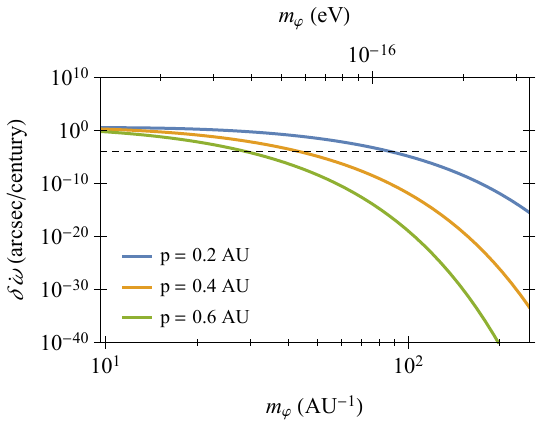}
\caption{Numerical solution for $\delta\dot\omega$. It was used $\omega_{0} = 100$, $e=0.5$ AU and $P = 87.97$ days for the orbit eccentricity and orbital period. The blue, yellow and green curves are constructed with $p=0.2$ AU, $p=0.4$ AU, and $p=0.6$ AU, respectively. The black dashed line is the uncertainty of the Mercury perihelion shift experiment \cite{Park_2017}.}
\label{fig3}
\end{figure}
\begin{figure}[t]
\includegraphics[width=\columnwidth]{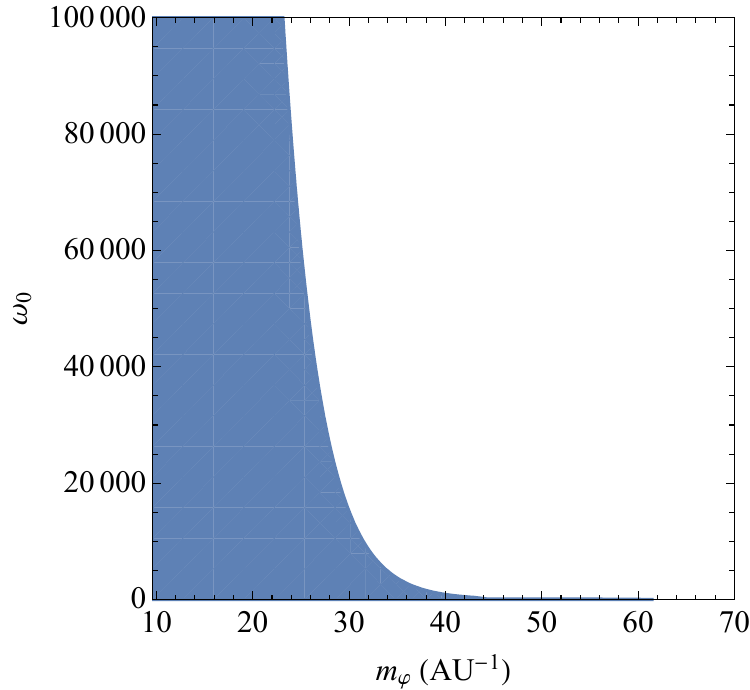}
\caption{The blue region shows the excluded parameter region due to Mercury perihelion advance. It was used $e=0.2056$ and $p=0.370$ AU for the orbit eccentricity and the \textit{semi-latus rectum}.}
\label{fig4}
\end{figure}

Due to the Yukawa correction (\ref{eq77}) we do not have a well-defined parameter $\beta$ since $\delta\dot\omega$ depends on the system constants $p$ and $e$ (see discussion in section \ref{sec:beta}). Despite this, we can use the fact that the best constraint in the $\beta$ parameter is about $10^{-5}$ \cite{Park_2017} and obtain constraints on the values of the parameters $\omega_0$ and $m_\varphi$. We can visualize these constraints by plotting the experimentally excluded region in parameter space. Figure \ref{fig4} shows the complete region of the parameter space, which is excluded by values of  $\delta\dot\omega \geq 10^ {-4}$ for the case of Mercury orbit ($p=0.370$ AU and $e = 0.2056$). The blue region shows for which parameter values we cannot define $\beta$ due to the Yukawa correction.  Outside this region, we can neglect the Yukawa correction, and consequently, we will get the same expression for $\dot\omega$ as in GR. In this case, it is possible to conclude that $\beta =1$, exactly as in GR.

In the next section, we will present another way to obtain a well-defined $\beta$ through an extended PPN version and the introduction of a new parameter that quantifies the dynamic effects associated with the Yukawa potential.

\section{The $\beta$ parameter and the Nordtvedt effect} \label{sec:beta}
In sections \ref{sec:pn} and \ref{sec:conserved} it was shown that fourth-order weighed Yukawa potential $\Phi_Y$ does not affect light motion neither the conserved post-Newtonian energy and momentum. Within the PPN formalism, these results demonstrate that the physical meaning associated to parameters  $\gamma$, $\zeta$'s and $\alpha$'s they are not changed by the presence of the potential $\Phi_Y$ outside the formalism. Once it was shown that photon geodesics and total energy and momentum expressions are the same as in GR, it was possible to conclude $\gamma=1$, as well as all $\zeta$'s and $\alpha$'s vanish.

The remaining PPN parameter to be determined, $\beta$, can be directly fixed by its influence on post-Newtonian periastron variations (since the others nine parameters are already fixed). This can be seen in the general PPN expression for $\Delta\omega$, namely
\begin{align}
    \Delta\omega_{PPN} = \ \frac{6\pi m}{p}&\left[\frac{1}{3}(2+2\gamma -\beta)\right. + \notag\\
    & \quad \left.\frac{1}{6}(2\alpha_1 -\alpha_2+\alpha_3 +2\zeta_2)\frac{\mu}{m}\right].\label{dw-ppn}
\end{align}
In the above expression, $\mu$ is the reduced mass of the binary system. Thus, one might find tempting, by comparing expressions \eqref{dw-ppn} and \eqref{periastronAdvance}, the definition of an effective $\beta$ parameter that would encompass the GR deviations brought by $\Phi_Y$. However, we understand that such treatment is not appropriated, once every PPN parameter must be considered strictly as a constant. If an effective parameter is defined as a function of coordinates or some system-dependent constants, one could not simply substitute the original parameter by its effective one in each observable physical phenomenon. For instance, if an effective $\beta$ were defined through \eqref{dw-ppn} and \eqref{periastronAdvance}, this new parameter would not have any influence on the so called Nordtvedt effect: the violation of the weak equivalence principle due to explicit contributions of self-gravitational energy to a body's inertial and gravitational mass.

The Nordtvedt effect can be tested in the Earth-Moon system by studying its motion in the Sun's gravitational field. If there is any difference between the free-fall acceleration of the Earth and Moon towards the Sun, this effect will be parametrized by the Nordtvedt parameter
\begin{equation}
\eta= 4\beta -\gamma -3 -\frac{10}{3}\,\xi -\alpha_1 +\frac{2}{3}\,\alpha_2 -\frac{2}{3}\,\zeta_1 -\frac{1}{3}\,\zeta_2,\label{nordtvedt}
\end{equation}
where $\xi$ is the PPN parameter related to the existence of preferred-locations effects.
It has been shown that $\Phi_Y$ does not bring any dependence in the acceleration of a body $A$ with its internal structure [cf. Eqs. \eqref{accN} -- \eqref{accStr}]. Consequently, no Nordtvedt effect is present and $\eta=0$, just like in GR. Therefore, a simple substitution of $\beta$, in \eqref{nordtvedt}, by an effective parameter obtained from the periastron advance would bring drastic and erroneous conclusions for any theory.

The correct approach to this issue would be to propose a new parameter, say $\tilde{\beta}$, to quantify the dynamical effects associated with $\Phi_Y$. In this extended PPN version, the Nordtvedt parameter will remain the same, while the periastron advance per orbit will be given by
\begin{align}
    \Delta\omega_{\rm EPPN}=\Delta\omega_{\rm PPN}+\tilde\beta\,\delta\omega.
\end{align}
In general relativity, for instance, $\beta=1$ and $\tilde\beta=0$, while in the case of a massive scalar-tensor theory with $\varphi\sim O(4)$ one has $\beta=1$ and $\tilde\beta=1$. Hence, with $\eta=0$ one also obtains $\xi=0$.

Thus, all the PPN parameters were determined as well as an extra parameter $\tilde\beta$. Their values are the same as in GR plus $\tilde\beta=1$. However, it is important to emphasize that, within this extended PPN approach, $\beta$ alone does not determine the post-Newtonian periastron advance effect. In this sense, the $\beta=1$ result is not enough to ensure agreement with Mercury's perihelion advance data, but it is determinant to guarantee no violations of the weak equivalence principle for fully conservative theories.

\section{CONCLUSION} \label{sec:conclusion}
We discussed the first-order post-Newtonian (PN) approximation of massive Brans-Dicke theories. This class of scalar-tensor theories spoils the Newtonian limit in general, adding an Yukawa-like contribution to the usual potential. Once the new gravitational potential is guided by the exponential of the scalar field effective mass ($m_\varphi$), two limiting situations arise. A small mass scalar field would approximate to the ordinary Brans-Dicke case, in which post-Newtonian behavior is well known to put a constraint of $\omega_0\gtrsim 4\times10^4$ to the theory's free parameter. On the other hand, a large mass can recover the Newtonian limit while shifting the Yukawa correction to the next order of approximation. We then focus on scenarios where the scalar field is considered to be of fourth order in the PN expansion.

The presence of this new potential hinders the direct determination of the PPN parameters from the metric. Consequently, we perform a derivation of all the conserved quantities and the acceleration of the center-of-mass bodies to verify whether and how $\Phi_Y$ does influence these effects. We show that conserved quantities are precisely the same as in GR, that is, the fourth-order Yukawa correction does not affect the amount of energy and momentum which is conserved. Meanwhile, the acceleration of massive bodies center of mass does have a non GR correction, leading to modifications in the orbital periastron advance. In addition to the well-known $10$ standard PPN parameters, we introduce and calculate a new parameter that captures the dynamics of the potential $\Phi_Y$, without point-particle approximations. The inclusion of this new parameter becomes essential to ensure consistency with the observed perihelion advance data of Mercury because the parameter $\beta$ alone is no longer adequate for explaining this phenomenon.

To quantify the extent of the modification caused by the new potential on the periastron advance, we introduce a correction term $\delta\dot\omega$ that depends on the orbital parameters of the system, as well as the scalar field effective mass and the theory's parameter $\omega_0$. We investigate the influence of this correction for different values of the theory's parameters and under variations of the orbital eccentricity and the semi-latus rectum. In particular, we analyze the behavior of $\delta\dot\omega$ in the orbits of Mercury and the S2 star orbiting Sgr $\text{A}^\ast$ in the center of our galaxy. Notwithstanding, the most strong observational constraint on perihelion shift comes from Mercury's orbit. Working with these data, we shown in Figure \ref{fig4} the necessary correlation between $\omega_0$ and $m_\varphi$ in order to obtain a massive Brans-Dicke model with $\varphi\sim O(4)$ in complete agreement with solar system tests.

In Ref. \cite{PhysRevD.85.064041} the authors obtain $\omega_0>40~000$ and the upper bound limit $m_\varphi\lesssim 2.5\times 10^{-20}~\text{eV}$, using data from Shapiro time delay once they worked with ordinary second-order scalar field (which also justify the presence of the Nordtvedt effect). Although they also work with the periastron shift effect the authors wrote final expressions for $\dot\omega$ only upon the massless and sufficiently large mass limits. On the other hand, our fourth-order weighed scalar field model allows to extract a lower bound limit of $m_\varphi\gtrsim 35 ~\text{AU}^{-1}$ or, equivalently, $m_\varphi\gtrsim 4.5\times10^{-15}~\text{eV}$.

To finish our discussion we comment on a specific method to extend the PPN (EPPN) formalism in order to include the fourth-order Yukawa potential. Instead of defining an effective $\beta$ parameter to deal with the Yukawa corrections to the perihelion shift, the EPPN would have a new parameter, $\tilde\beta$, apart from the original 10 PPN parameters. Thus $\beta=1$ would not be enough to guarantee agreement with Mercury's perihelion advance, it is also necessary to have $\tilde\beta=0$ (GR case) or, as in the case of massive scalar-tensor theories with $\varphi\sim O(4)$, $\tilde\beta\,\delta\dot\omega\lesssim 10^{-4}$. However, the $\beta=1$ value is necessary to non-violation of the weak equivalence principle, and this is independent of $\tilde\beta$.

\begin{acknowledgments}
MFSA thanks FAPES (Brazil) for its support. DCR thanks Heidelberg University for hospitality and support, and acknowledges support from \textit{Conselho Nacional de Desenvolvimento Científico e Tecnológico}  (CNPq - Brazil) and \textit{Fundação de Amparo à Pesquisa e Inovação do Espírito Santo} (FAPES - Brazil)  (1020/2022, 976/2022, 1081/2022).
\end{acknowledgments}

\bibliographystyle{apsrev4-1}

\bibliography{main}

\begin{thebibliography}{43}%
\makeatletter
\providecommand \@ifxundefined [1]{%
 \@ifx{#1\undefined}
}%
\providecommand \@ifnum [1]{%
 \ifnum #1\expandafter \@firstoftwo
 \else \expandafter \@secondoftwo
 \fi
}%
\providecommand \@ifx [1]{%
 \ifx #1\expandafter \@firstoftwo
 \else \expandafter \@secondoftwo
 \fi
}%
\providecommand \natexlab [1]{#1}%
\providecommand \enquote  [1]{``#1''}%
\providecommand \bibnamefont  [1]{#1}%
\providecommand \bibfnamefont [1]{#1}%
\providecommand \citenamefont [1]{#1}%
\providecommand \href@noop [0]{\@secondoftwo}%
\providecommand \href [0]{\begingroup \@sanitize@url \@href}%
\providecommand \@href[1]{\@@startlink{#1}\@@href}%
\providecommand \@@href[1]{\endgroup#1\@@endlink}%
\providecommand \@sanitize@url [0]{\catcode `\\12\catcode `\$12\catcode `\&12\catcode `\#12\catcode `\^12\catcode `\_12\catcode `\%12\relax}%
\providecommand \@@startlink[1]{}%
\providecommand \@@endlink[0]{}%
\providecommand \url  [0]{\begingroup\@sanitize@url \@url }%
\providecommand \@url [1]{\endgroup\@href {#1}{\urlprefix }}%
\providecommand \urlprefix  [0]{URL }%
\providecommand \Eprint [0]{\href }%
\providecommand \doibase [0]{http://dx.doi.org/}%
\providecommand \selectlanguage [0]{\@gobble}%
\providecommand \bibinfo  [0]{\@secondoftwo}%
\providecommand \bibfield  [0]{\@secondoftwo}%
\providecommand \translation [1]{[#1]}%
\providecommand \BibitemOpen [0]{}%
\providecommand \bibitemStop [0]{}%
\providecommand \bibitemNoStop [0]{.\EOS\space}%
\providecommand \EOS [0]{\spacefactor3000\relax}%
\providecommand \BibitemShut  [1]{\csname bibitem#1\endcsname}%
\let\auto@bib@innerbib\@empty
\bibitem [{\citenamefont {Brans}\ and\ \citenamefont {Dicke}(1961)}]{PhysRev.124.925}%
  \BibitemOpen
  \bibfield  {author} {\bibinfo {author} {\bibfnamefont {C.}~\bibnamefont {Brans}}\ and\ \bibinfo {author} {\bibfnamefont {R.~H.}\ \bibnamefont {Dicke}},\ }\href {\doibase 10.1103/PhysRev.124.925} {\bibfield  {journal} {\bibinfo  {journal} {Phys. Rev.}\ }\textbf {\bibinfo {volume} {124}},\ \bibinfo {pages} {925} (\bibinfo {year} {1961})}\BibitemShut {NoStop}%
\bibitem [{\citenamefont {Bergmann}(1968)}]{bergmann1968comments}%
  \BibitemOpen
  \bibfield  {author} {\bibinfo {author} {\bibfnamefont {P.~G.}\ \bibnamefont {Bergmann}},\ }\href@noop {} {\bibfield  {journal} {\bibinfo  {journal} {International Journal of Theoretical Physics}\ }\textbf {\bibinfo {volume} {1}},\ \bibinfo {pages} {25} (\bibinfo {year} {1968})}\BibitemShut {NoStop}%
\bibitem [{\citenamefont {Wagoner}(1970)}]{wagoner1970scalar}%
  \BibitemOpen
  \bibfield  {author} {\bibinfo {author} {\bibfnamefont {R.~V.}\ \bibnamefont {Wagoner}},\ }\href@noop {} {\bibfield  {journal} {\bibinfo  {journal} {Physical Review D}\ }\textbf {\bibinfo {volume} {1}},\ \bibinfo {pages} {3209} (\bibinfo {year} {1970})}\BibitemShut {NoStop}%
\bibitem [{\citenamefont {Nordtvedt~Jr}(1970)}]{nordtvedt1970post}%
  \BibitemOpen
  \bibfield  {author} {\bibinfo {author} {\bibfnamefont {K.}~\bibnamefont {Nordtvedt~Jr}},\ }\href@noop {} {\bibfield  {journal} {\bibinfo  {journal} {The Astrophysical Journal}\ }\textbf {\bibinfo {volume} {161}},\ \bibinfo {pages} {1059} (\bibinfo {year} {1970})}\BibitemShut {NoStop}%
\bibitem [{\citenamefont {Damour}\ and\ \citenamefont {Esposito-Farese}(1992)}]{damour1992tensor}%
  \BibitemOpen
  \bibfield  {author} {\bibinfo {author} {\bibfnamefont {T.}~\bibnamefont {Damour}}\ and\ \bibinfo {author} {\bibfnamefont {G.}~\bibnamefont {Esposito-Farese}},\ }\href@noop {} {\bibfield  {journal} {\bibinfo  {journal} {Classical and Quantum Gravity}\ }\textbf {\bibinfo {volume} {9}},\ \bibinfo {pages} {2093} (\bibinfo {year} {1992})}\BibitemShut {NoStop}%
\bibitem [{\citenamefont {Fujii}\ and\ \citenamefont {Maeda}(2003)}]{fujii2003scalar}%
  \BibitemOpen
  \bibfield  {author} {\bibinfo {author} {\bibfnamefont {Y.}~\bibnamefont {Fujii}}\ and\ \bibinfo {author} {\bibfnamefont {K.-i.}\ \bibnamefont {Maeda}},\ }\href@noop {} {\emph {\bibinfo {title} {The scalar-tensor theory of gravitation}}}\ (\bibinfo  {publisher} {Cambridge University Press},\ \bibinfo {year} {2003})\BibitemShut {NoStop}%
\bibitem [{\citenamefont {Faraoni}\ and\ \citenamefont {Faraoni}(2004)}]{faraoni2004scalar}%
  \BibitemOpen
  \bibfield  {author} {\bibinfo {author} {\bibfnamefont {V.}~\bibnamefont {Faraoni}}\ and\ \bibinfo {author} {\bibfnamefont {V.}~\bibnamefont {Faraoni}},\ }\href@noop {} {\emph {\bibinfo {title} {Scalar-Tensor Gravity}}}\ (\bibinfo  {publisher} {Springer},\ \bibinfo {year} {2004})\BibitemShut {NoStop}%
\bibitem [{\citenamefont {Esposito-Far\`ese}\ and\ \citenamefont {Polarski}(2001)}]{PhysRevD.63.063504}%
  \BibitemOpen
  \bibfield  {author} {\bibinfo {author} {\bibfnamefont {G.}~\bibnamefont {Esposito-Far\`ese}}\ and\ \bibinfo {author} {\bibfnamefont {D.}~\bibnamefont {Polarski}},\ }\href {\doibase 10.1103/PhysRevD.63.063504} {\bibfield  {journal} {\bibinfo  {journal} {Phys. Rev. D}\ }\textbf {\bibinfo {volume} {63}},\ \bibinfo {pages} {063504} (\bibinfo {year} {2001})}\BibitemShut {NoStop}%
\bibitem [{\citenamefont {Éanna É~Flanagan}(2004)}]{EFlanagan_2004}%
  \BibitemOpen
  \bibfield  {author} {\bibinfo {author} {\bibnamefont {Éanna É~Flanagan}},\ }\href {\doibase 10.1088/0264-9381/21/15/N02} {\bibfield  {journal} {\bibinfo  {journal} {Classical and Quantum Gravity}\ }\textbf {\bibinfo {volume} {21}},\ \bibinfo {pages} {3817} (\bibinfo {year} {2004})}\BibitemShut {NoStop}%
\bibitem [{\citenamefont {Clifton}\ \emph {et~al.}(2012)\citenamefont {Clifton}, \citenamefont {Ferreira}, \citenamefont {Padilla},\ and\ \citenamefont {Skordis}}]{clifton2012modified}%
  \BibitemOpen
  \bibfield  {author} {\bibinfo {author} {\bibfnamefont {T.}~\bibnamefont {Clifton}}, \bibinfo {author} {\bibfnamefont {P.~G.}\ \bibnamefont {Ferreira}}, \bibinfo {author} {\bibfnamefont {A.}~\bibnamefont {Padilla}}, \ and\ \bibinfo {author} {\bibfnamefont {C.}~\bibnamefont {Skordis}},\ }\href@noop {} {\bibfield  {journal} {\bibinfo  {journal} {Physics reports}\ }\textbf {\bibinfo {volume} {513}},\ \bibinfo {pages} {1} (\bibinfo {year} {2012})}\BibitemShut {NoStop}%
\bibitem [{\citenamefont {Santos}\ and\ \citenamefont {Gregory}(1997)}]{santos1997cosmology}%
  \BibitemOpen
  \bibfield  {author} {\bibinfo {author} {\bibfnamefont {C.}~\bibnamefont {Santos}}\ and\ \bibinfo {author} {\bibfnamefont {R.}~\bibnamefont {Gregory}},\ }\href@noop {} {\bibfield  {journal} {\bibinfo  {journal} {Annals of Physics}\ }\textbf {\bibinfo {volume} {258}},\ \bibinfo {pages} {111} (\bibinfo {year} {1997})}\BibitemShut {NoStop}%
\bibitem [{\citenamefont {Toniato}\ and\ \citenamefont {Rodrigues}(2021)}]{PhysRevD.104.044020}%
  \BibitemOpen
  \bibfield  {author} {\bibinfo {author} {\bibfnamefont {J.~D.}\ \bibnamefont {Toniato}}\ and\ \bibinfo {author} {\bibfnamefont {D.~C.}\ \bibnamefont {Rodrigues}},\ }\href {\doibase 10.1103/PhysRevD.104.044020} {\bibfield  {journal} {\bibinfo  {journal} {Phys. Rev. D}\ }\textbf {\bibinfo {volume} {104}},\ \bibinfo {pages} {044020} (\bibinfo {year} {2021})}\BibitemShut {NoStop}%
\bibitem [{\citenamefont {Carmichael}(1925)}]{carmichael1925eddington}%
  \BibitemOpen
  \bibfield  {author} {\bibinfo {author} {\bibfnamefont {R.}~\bibnamefont {Carmichael}},\ }\href@noop {} {\  (\bibinfo {year} {1925})}\BibitemShut {NoStop}%
\bibitem [{\citenamefont {Robertson}(1962)}]{robertson1962relativity}%
  \BibitemOpen
  \bibfield  {author} {\bibinfo {author} {\bibfnamefont {H.~P.}\ \bibnamefont {Robertson}},\ }in\ \href@noop {} {\emph {\bibinfo {booktitle} {Space age astronomy}}}\ (\bibinfo {year} {1962})\ p.\ \bibinfo {pages} {228}\BibitemShut {NoStop}%
\bibitem [{\citenamefont {Schiff}(1966)}]{schiff1966comparison}%
  \BibitemOpen
  \bibfield  {author} {\bibinfo {author} {\bibfnamefont {L.}~\bibnamefont {Schiff}},\ }\href@noop {} {\emph {\bibinfo {title} {Comparison of theory and observation in general relativity}}},\ \bibinfo {type} {Tech. Rep.}\ (\bibinfo  {institution} {INSTITUTE OF THEORETICAL PHYSICS STANFORD UNIV CALIF},\ \bibinfo {year} {1966})\BibitemShut {NoStop}%
\bibitem [{\citenamefont {Nordtvedt}(1968)}]{PhysRev.169.1017}%
  \BibitemOpen
  \bibfield  {author} {\bibinfo {author} {\bibfnamefont {K.}~\bibnamefont {Nordtvedt}},\ }\href {\doibase 10.1103/PhysRev.169.1017} {\bibfield  {journal} {\bibinfo  {journal} {Phys. Rev.}\ }\textbf {\bibinfo {volume} {169}},\ \bibinfo {pages} {1017} (\bibinfo {year} {1968})}\BibitemShut {NoStop}%
\bibitem [{\citenamefont {Poisson}\ and\ \citenamefont {Will}(2014)}]{poisson_will_2014}%
  \BibitemOpen
  \bibfield  {author} {\bibinfo {author} {\bibfnamefont {E.}~\bibnamefont {Poisson}}\ and\ \bibinfo {author} {\bibfnamefont {C.~M.}\ \bibnamefont {Will}},\ }\href {\doibase 10.1017/CBO9781139507486} {\emph {\bibinfo {title} {Gravity: Newtonian, Post-Newtonian, Relativistic}}}\ (\bibinfo  {publisher} {Cambridge University Press},\ \bibinfo {year} {2014})\BibitemShut {NoStop}%
\bibitem [{\citenamefont {Will}(2014)}]{Will_2014}%
  \BibitemOpen
  \bibfield  {author} {\bibinfo {author} {\bibfnamefont {C.~M.}\ \bibnamefont {Will}},\ }\href {\doibase 10.12942/lrr-2014-4} {\bibfield  {journal} {\bibinfo  {journal} {Living Reviews in Relativity}\ }\textbf {\bibinfo {volume} {17}} (\bibinfo {year} {2014}),\ 10.12942/lrr-2014-4}\BibitemShut {NoStop}%
\bibitem [{\citenamefont {Will}(2018)}]{will_2018}%
  \BibitemOpen
  \bibfield  {author} {\bibinfo {author} {\bibfnamefont {C.~M.}\ \bibnamefont {Will}},\ }\href {\doibase 10.1017/9781316338612} {\emph {\bibinfo {title} {Theory and Experiment in Gravitational Physics}}},\ \bibinfo {edition} {2nd}\ ed.\ (\bibinfo  {publisher} {Cambridge University Press},\ \bibinfo {year} {2018})\BibitemShut {NoStop}%
\bibitem [{\citenamefont {Fomalont}\ \emph {et~al.}(2009)\citenamefont {Fomalont}, \citenamefont {Kopeikin}, \citenamefont {Lanyi},\ and\ \citenamefont {Benson}}]{fomalont2009progress}%
  \BibitemOpen
  \bibfield  {author} {\bibinfo {author} {\bibfnamefont {E.}~\bibnamefont {Fomalont}}, \bibinfo {author} {\bibfnamefont {S.}~\bibnamefont {Kopeikin}}, \bibinfo {author} {\bibfnamefont {G.}~\bibnamefont {Lanyi}}, \ and\ \bibinfo {author} {\bibfnamefont {J.}~\bibnamefont {Benson}},\ }\href@noop {} {\bibfield  {journal} {\bibinfo  {journal} {The Astrophysical Journal}\ }\textbf {\bibinfo {volume} {699}},\ \bibinfo {pages} {1395} (\bibinfo {year} {2009})}\BibitemShut {NoStop}%
\bibitem [{\citenamefont {Bertotti}\ \emph {et~al.}(2003)\citenamefont {Bertotti}, \citenamefont {Iess},\ and\ \citenamefont {Tortora}}]{bertotti2003test}%
  \BibitemOpen
  \bibfield  {author} {\bibinfo {author} {\bibfnamefont {B.}~\bibnamefont {Bertotti}}, \bibinfo {author} {\bibfnamefont {L.}~\bibnamefont {Iess}}, \ and\ \bibinfo {author} {\bibfnamefont {P.}~\bibnamefont {Tortora}},\ }\href@noop {} {\bibfield  {journal} {\bibinfo  {journal} {Nature}\ }\textbf {\bibinfo {volume} {425}},\ \bibinfo {pages} {374} (\bibinfo {year} {2003})}\BibitemShut {NoStop}%
\bibitem [{\citenamefont {Hofmann}\ \emph {et~al.}(2010)\citenamefont {Hofmann}, \citenamefont {M{\"u}ller},\ and\ \citenamefont {Biskupek}}]{hofmann2010lunar}%
  \BibitemOpen
  \bibfield  {author} {\bibinfo {author} {\bibfnamefont {F.}~\bibnamefont {Hofmann}}, \bibinfo {author} {\bibfnamefont {J.}~\bibnamefont {M{\"u}ller}}, \ and\ \bibinfo {author} {\bibfnamefont {L.}~\bibnamefont {Biskupek}},\ }\href@noop {} {\bibfield  {journal} {\bibinfo  {journal} {Astronomy \& Astrophysics}\ }\textbf {\bibinfo {volume} {522}},\ \bibinfo {pages} {L5} (\bibinfo {year} {2010})}\BibitemShut {NoStop}%
\bibitem [{\citenamefont {Lambert}\ and\ \citenamefont {Le~Poncin-Lafitte}(2011)}]{lambert2011improved}%
  \BibitemOpen
  \bibfield  {author} {\bibinfo {author} {\bibfnamefont {S.}~\bibnamefont {Lambert}}\ and\ \bibinfo {author} {\bibfnamefont {C.}~\bibnamefont {Le~Poncin-Lafitte}},\ }\href@noop {} {\bibfield  {journal} {\bibinfo  {journal} {Astronomy \& Astrophysics}\ }\textbf {\bibinfo {volume} {529}},\ \bibinfo {pages} {A70} (\bibinfo {year} {2011})}\BibitemShut {NoStop}%
\bibitem [{\citenamefont {Fienga}\ \emph {et~al.}(2011)\citenamefont {Fienga}, \citenamefont {Laskar}, \citenamefont {Kuchynka}, \citenamefont {Manche}, \citenamefont {Desvignes}, \citenamefont {Gastineau}, \citenamefont {Cognard},\ and\ \citenamefont {Theureau}}]{fienga2011inpop10a}%
  \BibitemOpen
  \bibfield  {author} {\bibinfo {author} {\bibfnamefont {A.}~\bibnamefont {Fienga}}, \bibinfo {author} {\bibfnamefont {J.}~\bibnamefont {Laskar}}, \bibinfo {author} {\bibfnamefont {P.}~\bibnamefont {Kuchynka}}, \bibinfo {author} {\bibfnamefont {H.}~\bibnamefont {Manche}}, \bibinfo {author} {\bibfnamefont {G.}~\bibnamefont {Desvignes}}, \bibinfo {author} {\bibfnamefont {M.}~\bibnamefont {Gastineau}}, \bibinfo {author} {\bibfnamefont {I.}~\bibnamefont {Cognard}}, \ and\ \bibinfo {author} {\bibfnamefont {G.}~\bibnamefont {Theureau}},\ }\href@noop {} {\bibfield  {journal} {\bibinfo  {journal} {Celestial Mechanics and Dynamical Astronomy}\ }\textbf {\bibinfo {volume} {111}},\ \bibinfo {pages} {363} (\bibinfo {year} {2011})}\BibitemShut {NoStop}%
\bibitem [{\citenamefont {Pitjeva}\ and\ \citenamefont {Pitjev}(2013)}]{pitjeva2013relativistic}%
  \BibitemOpen
  \bibfield  {author} {\bibinfo {author} {\bibfnamefont {E.}~\bibnamefont {Pitjeva}}\ and\ \bibinfo {author} {\bibfnamefont {N.}~\bibnamefont {Pitjev}},\ }\href@noop {} {\bibfield  {journal} {\bibinfo  {journal} {Monthly Notices of the Royal Astronomical Society}\ }\textbf {\bibinfo {volume} {432}},\ \bibinfo {pages} {3431} (\bibinfo {year} {2013})}\BibitemShut {NoStop}%
\bibitem [{\citenamefont {Park}\ \emph {et~al.}(2017)\citenamefont {Park}, \citenamefont {Folkner}, \citenamefont {Konopliv}, \citenamefont {Williams}, \citenamefont {Smith},\ and\ \citenamefont {Zuber}}]{Park_2017}%
  \BibitemOpen
  \bibfield  {author} {\bibinfo {author} {\bibfnamefont {R.~S.}\ \bibnamefont {Park}}, \bibinfo {author} {\bibfnamefont {W.~M.}\ \bibnamefont {Folkner}}, \bibinfo {author} {\bibfnamefont {A.~S.}\ \bibnamefont {Konopliv}}, \bibinfo {author} {\bibfnamefont {J.~G.}\ \bibnamefont {Williams}}, \bibinfo {author} {\bibfnamefont {D.~E.}\ \bibnamefont {Smith}}, \ and\ \bibinfo {author} {\bibfnamefont {M.~T.}\ \bibnamefont {Zuber}},\ }\href {\doibase 10.3847/1538-3881/aa5be2} {\bibfield  {journal} {\bibinfo  {journal} {The Astronomical Journal}\ }\textbf {\bibinfo {volume} {153}},\ \bibinfo {pages} {121} (\bibinfo {year} {2017})}\BibitemShut {NoStop}%
\bibitem [{\citenamefont {Perivolaropoulos}(2010)}]{PhysRevD.81.047501}%
  \BibitemOpen
  \bibfield  {author} {\bibinfo {author} {\bibfnamefont {L.}~\bibnamefont {Perivolaropoulos}},\ }\href {\doibase 10.1103/PhysRevD.81.047501} {\bibfield  {journal} {\bibinfo  {journal} {Phys. Rev. D}\ }\textbf {\bibinfo {volume} {81}},\ \bibinfo {pages} {047501} (\bibinfo {year} {2010})}\BibitemShut {NoStop}%
\bibitem [{\citenamefont {Hohmann}\ \emph {et~al.}(2016)\citenamefont {Hohmann}, \citenamefont {J\"arv}, \citenamefont {Kuusk}, \citenamefont {Randla},\ and\ \citenamefont {Vilson}}]{PhysRevD.94.124015}%
  \BibitemOpen
  \bibfield  {author} {\bibinfo {author} {\bibfnamefont {M.}~\bibnamefont {Hohmann}}, \bibinfo {author} {\bibfnamefont {L.}~\bibnamefont {J\"arv}}, \bibinfo {author} {\bibfnamefont {P.}~\bibnamefont {Kuusk}}, \bibinfo {author} {\bibfnamefont {E.}~\bibnamefont {Randla}}, \ and\ \bibinfo {author} {\bibfnamefont {O.}~\bibnamefont {Vilson}},\ }\href {\doibase 10.1103/PhysRevD.94.124015} {\bibfield  {journal} {\bibinfo  {journal} {Phys. Rev. D}\ }\textbf {\bibinfo {volume} {94}},\ \bibinfo {pages} {124015} (\bibinfo {year} {2016})}\BibitemShut {NoStop}%
\bibitem [{\citenamefont {Hohmann}\ \emph {et~al.}(2013)\citenamefont {Hohmann}, \citenamefont {J\"arv}, \citenamefont {Kuusk},\ and\ \citenamefont {Randla}}]{PhysRevD.88.084054}%
  \BibitemOpen
  \bibfield  {author} {\bibinfo {author} {\bibfnamefont {M.}~\bibnamefont {Hohmann}}, \bibinfo {author} {\bibfnamefont {L.}~\bibnamefont {J\"arv}}, \bibinfo {author} {\bibfnamefont {P.}~\bibnamefont {Kuusk}}, \ and\ \bibinfo {author} {\bibfnamefont {E.}~\bibnamefont {Randla}},\ }\href {\doibase 10.1103/PhysRevD.88.084054} {\bibfield  {journal} {\bibinfo  {journal} {Phys. Rev. D}\ }\textbf {\bibinfo {volume} {88}},\ \bibinfo {pages} {084054} (\bibinfo {year} {2013})}\BibitemShut {NoStop}%
\bibitem [{\citenamefont {Toniato}\ \emph {et~al.}(2020)\citenamefont {Toniato}, \citenamefont {Rodrigues},\ and\ \citenamefont {Wojnar}}]{PhysRevD.101.064050}%
  \BibitemOpen
  \bibfield  {author} {\bibinfo {author} {\bibfnamefont {J.~D.}\ \bibnamefont {Toniato}}, \bibinfo {author} {\bibfnamefont {D.~C.}\ \bibnamefont {Rodrigues}}, \ and\ \bibinfo {author} {\bibfnamefont {A.}~\bibnamefont {Wojnar}},\ }\href {\doibase 10.1103/PhysRevD.101.064050} {\bibfield  {journal} {\bibinfo  {journal} {Phys. Rev. D}\ }\textbf {\bibinfo {volume} {101}},\ \bibinfo {pages} {064050} (\bibinfo {year} {2020})}\BibitemShut {NoStop}%
\bibitem [{\citenamefont {Capone}\ and\ \citenamefont {Ruggiero}(2010)}]{Capone_2010}%
  \BibitemOpen
  \bibfield  {author} {\bibinfo {author} {\bibfnamefont {M.}~\bibnamefont {Capone}}\ and\ \bibinfo {author} {\bibfnamefont {M.~L.}\ \bibnamefont {Ruggiero}},\ }\href {\doibase 10.1088/0264-9381/27/12/125006} {\bibfield  {journal} {\bibinfo  {journal} {Classical and Quantum Gravity}\ }\textbf {\bibinfo {volume} {27}},\ \bibinfo {pages} {125006} (\bibinfo {year} {2010})}\BibitemShut {NoStop}%
\bibitem [{\citenamefont {Chiba}(2003)}]{chiba20031}%
  \BibitemOpen
  \bibfield  {author} {\bibinfo {author} {\bibfnamefont {T.}~\bibnamefont {Chiba}},\ }\href@noop {} {\bibfield  {journal} {\bibinfo  {journal} {Physics Letters B}\ }\textbf {\bibinfo {volume} {575}},\ \bibinfo {pages} {1} (\bibinfo {year} {2003})}\BibitemShut {NoStop}%
\bibitem [{\citenamefont {Faraoni}(2007)}]{faraoni2007sitter}%
  \BibitemOpen
  \bibfield  {author} {\bibinfo {author} {\bibfnamefont {V.}~\bibnamefont {Faraoni}},\ }\href@noop {} {\bibfield  {journal} {\bibinfo  {journal} {Physical Review D}\ }\textbf {\bibinfo {volume} {75}},\ \bibinfo {pages} {067302} (\bibinfo {year} {2007})}\BibitemShut {NoStop}%
\bibitem [{\citenamefont {Flanagan}(2004)}]{flanagan2004palatini}%
  \BibitemOpen
  \bibfield  {author} {\bibinfo {author} {\bibfnamefont {E.~E.}\ \bibnamefont {Flanagan}},\ }\href@noop {} {\bibfield  {journal} {\bibinfo  {journal} {Physical review letters}\ }\textbf {\bibinfo {volume} {92}},\ \bibinfo {pages} {071101} (\bibinfo {year} {2004})}\BibitemShut {NoStop}%
\bibitem [{\citenamefont {Sotiriou}\ and\ \citenamefont {Faraoni}(2010)}]{RevModPhys.82.451}%
  \BibitemOpen
  \bibfield  {author} {\bibinfo {author} {\bibfnamefont {T.~P.}\ \bibnamefont {Sotiriou}}\ and\ \bibinfo {author} {\bibfnamefont {V.}~\bibnamefont {Faraoni}},\ }\href {\doibase 10.1103/RevModPhys.82.451} {\bibfield  {journal} {\bibinfo  {journal} {Rev. Mod. Phys.}\ }\textbf {\bibinfo {volume} {82}},\ \bibinfo {pages} {451} (\bibinfo {year} {2010})}\BibitemShut {NoStop}%
\bibitem [{\citenamefont {Nutku}(1969)}]{nutku1969post}%
  \BibitemOpen
  \bibfield  {author} {\bibinfo {author} {\bibfnamefont {Y.}~\bibnamefont {Nutku}},\ }\href@noop {} {\bibfield  {journal} {\bibinfo  {journal} {Astrophysical Journal, vol. 155, p. 999}\ }\textbf {\bibinfo {volume} {155}},\ \bibinfo {pages} {999} (\bibinfo {year} {1969})}\BibitemShut {NoStop}%
\bibitem [{\citenamefont {Klimek}(2009)}]{klimek2009parameterized}%
  \BibitemOpen
  \bibfield  {author} {\bibinfo {author} {\bibfnamefont {M.~D.}\ \bibnamefont {Klimek}},\ }\href@noop {} {\bibfield  {journal} {\bibinfo  {journal} {Classical and Quantum Gravity}\ }\textbf {\bibinfo {volume} {26}},\ \bibinfo {pages} {065005} (\bibinfo {year} {2009})}\BibitemShut {NoStop}%
\bibitem [{\citenamefont {Alsing}\ \emph {et~al.}(2012)\citenamefont {Alsing}, \citenamefont {Berti}, \citenamefont {Will},\ and\ \citenamefont {Zaglauer}}]{PhysRevD.85.064041}%
  \BibitemOpen
  \bibfield  {author} {\bibinfo {author} {\bibfnamefont {J.}~\bibnamefont {Alsing}}, \bibinfo {author} {\bibfnamefont {E.}~\bibnamefont {Berti}}, \bibinfo {author} {\bibfnamefont {C.~M.}\ \bibnamefont {Will}}, \ and\ \bibinfo {author} {\bibfnamefont {H.}~\bibnamefont {Zaglauer}},\ }\href {\doibase 10.1103/PhysRevD.85.064041} {\bibfield  {journal} {\bibinfo  {journal} {Phys. Rev. D}\ }\textbf {\bibinfo {volume} {85}},\ \bibinfo {pages} {064041} (\bibinfo {year} {2012})}\BibitemShut {NoStop}%
\bibitem [{\citenamefont {Faraoni}\ \emph {et~al.}(1998)\citenamefont {Faraoni}, \citenamefont {Gunzig},\ and\ \citenamefont {Nardone}}]{faraoni1998conformal}%
  \BibitemOpen
  \bibfield  {author} {\bibinfo {author} {\bibfnamefont {V.}~\bibnamefont {Faraoni}}, \bibinfo {author} {\bibfnamefont {E.}~\bibnamefont {Gunzig}}, \ and\ \bibinfo {author} {\bibfnamefont {P.}~\bibnamefont {Nardone}},\ }\href@noop {} {\enquote {\bibinfo {title} {Conformal transformations in classical gravitational theories and in cosmology},}\ } (\bibinfo {year} {1998}),\ \Eprint {http://arxiv.org/abs/gr-qc/9811047} {arXiv:gr-qc/9811047 [gr-qc]} \BibitemShut {NoStop}%
\bibitem [{\citenamefont {Sotiriou}\ and\ \citenamefont {Faraoni}(2012)}]{Sotiriou:2011dz}%
  \BibitemOpen
  \bibfield  {author} {\bibinfo {author} {\bibfnamefont {T.~P.}\ \bibnamefont {Sotiriou}}\ and\ \bibinfo {author} {\bibfnamefont {V.}~\bibnamefont {Faraoni}},\ }\href {\doibase 10.1103/PhysRevLett.108.081103} {\bibfield  {journal} {\bibinfo  {journal} {Phys. Rev. Lett.}\ }\textbf {\bibinfo {volume} {108}},\ \bibinfo {pages} {081103} (\bibinfo {year} {2012})},\ \Eprint {http://arxiv.org/abs/1109.6324} {arXiv:1109.6324 [gr-qc]} \BibitemShut {NoStop}%
\bibitem [{\citenamefont {Abuter}\ \emph {et~al.}(2020)\citenamefont {Abuter} \emph {et~al.}}]{GRAVITY:2020gka}%
  \BibitemOpen
  \bibfield  {author} {\bibinfo {author} {\bibfnamefont {R.}~\bibnamefont {Abuter}} \emph {et~al.} (\bibinfo {collaboration} {GRAVITY}),\ }\href {\doibase 10.1051/0004-6361/202037813} {\bibfield  {journal} {\bibinfo  {journal} {Astron. Astrophys.}\ }\textbf {\bibinfo {volume} {636}},\ \bibinfo {pages} {L5} (\bibinfo {year} {2020})},\ \Eprint {http://arxiv.org/abs/2004.07187} {arXiv:2004.07187 [astro-ph.GA]} \BibitemShut {NoStop}%
\bibitem [{\citenamefont {Merritt}\ \emph {et~al.}(2010)\citenamefont {Merritt}, \citenamefont {Alexander}, \citenamefont {Mikkola},\ and\ \citenamefont {Will}}]{Merritt_2010}%
  \BibitemOpen
  \bibfield  {author} {\bibinfo {author} {\bibfnamefont {D.}~\bibnamefont {Merritt}}, \bibinfo {author} {\bibfnamefont {T.}~\bibnamefont {Alexander}}, \bibinfo {author} {\bibfnamefont {S.}~\bibnamefont {Mikkola}}, \ and\ \bibinfo {author} {\bibfnamefont {C.~M.}\ \bibnamefont {Will}},\ }\href {\doibase 10.1103/physrevd.81.062002} {\bibfield  {journal} {\bibinfo  {journal} {Physical Review D}\ }\textbf {\bibinfo {volume} {81}} (\bibinfo {year} {2010}),\ 10.1103/physrevd.81.062002}\BibitemShut {NoStop}%
\bibitem [{\citenamefont {Zhang}\ and\ \citenamefont {Iorio}(2017)}]{Zhang_2017}%
  \BibitemOpen
  \bibfield  {author} {\bibinfo {author} {\bibfnamefont {F.}~\bibnamefont {Zhang}}\ and\ \bibinfo {author} {\bibfnamefont {L.}~\bibnamefont {Iorio}},\ }\href {\doibase 10.3847/1538-4357/834/2/198} {\bibfield  {journal} {\bibinfo  {journal} {The Astrophysical Journal}\ }\textbf {\bibinfo {volume} {834}},\ \bibinfo {pages} {198} (\bibinfo {year} {2017})}\BibitemShut {NoStop}%
\end{thebibliography}%
\end{document}